\documentclass[twocolumn]{jpsj3}
\usepackage{graphicx}
\usepackage{amsmath,amssymb}

\def\runauthor{Youichi {\sc Yanase}}

\title{ 
Antiferromagnetic Phases in the
Fulde-Ferrell-Larkin-Ovchinnikov State of CeCoIn$_5$ 
} 

\author{Youichi {\sc Yanase}$^{1,2}$\footnote{E-mail:
yanase@phys.sc.niigata-u.ac.jp} and Manfred Sigrist$^2$}

\inst{
$^{1}$Department of Physics, Niigata University, Ikarashi, 
Niigata 950-2181, Japan
\\
$^2$Theoretische Physik, ETH-Honggerberg, 8093 Zurich, Switzerland
}

\recdate{Today 2010}

\abst
{ 
 The antiferromagnetic (AFM) order in the Fulde-Ferrell-Larkin-Ovchinnikov 
(FFLO) superconducting state is analyzed on the basis of a
Ginzburg-Landau theory. 
 To examine the possible AFM-FFLO state in CeCoIn$_5$, 
we focus on the incommensurate AFM order characterized by the 
wave vector $\vec{Q} = \vec{Q}_{0} \pm \vec{q}_{\rm inc}$ with 
$\vec{Q}_{0} =(\pi,\pi,\pi) $ and 
$\vec{q}_{\rm inc} \parallel [110]$ or $[1\bar{1}0]$
in the tetragonal crystal structure. 
 We formulate the two component Ginzburg-Landau theory 
and investigate the two degenerate incommensurate AFM order with 
$\vec{q}_{\rm inc} \parallel [110]$ and $[1\bar{1}0]$.
 We show that the pinning of AFM moment due to the FFLO nodal planes 
leads to multiple phases in magnetic fields along
$ [100]$ or $[010]$. 
 The phase diagrams for various coupling constants between 
the two order parameters are shown for the comparison with CeCoIn$_5$. 
 Experimental results of the NMR and neutron scattering measurements 
are discussed. 
}

\kword
{
FFLO superconductivity, unconventional magnetism, CeCoIn$_5$ 
}

\begin{document}
\sloppy
\maketitle

\newcommand{\eli}{$\acute{{\rm E}}$liashberg }
\renewcommand{\k}{\vec{k}}
\newcommand{\kk}{\vec{k'}}
\newcommand{\q}{\vec{q}}
\newcommand{\qa}{\vec{q}_{1}}
\newcommand{\qb}{\vec{q}_{2}}
\newcommand{\Q}{\vec{Q}}
\newcommand{\Qaf}{\vec{Q}_{0}}
\newcommand{\qi}{q_{\rm inc}}
\newcommand{\qiv}{\vec{q}_{\rm inc}}
\newcommand{\qia}{\vec{q}_{\rm inc}^{\rm \,\,(1)}}
\newcommand{\qib}{\vec{q}_{\rm inc}^{\rm \,\,(2)}}
\renewcommand{\r}{\vec{r}}
\newcommand{\e}{\varepsilon}
\newcommand{\ee}{\varepsilon^{'}}
\newcommand{\ep}{\varepsilon}
\newcommand{\s}{{\mit{\it \Sigma}}}
\newcommand{\Tc}{$T_{\rm c}$ }
\newcommand{\Tcf}{$T_{\rm c}$}
\newcommand{\TN}{$T_{\rm N}$ }
\newcommand{\TNz}{$T_{\rm N}^0$ }
\newcommand{\TNf}{$T_{\rm N}$}
\newcommand{\Hc}{$H_{\rm c2}^{\rm P}$ }
\newcommand{\Hcf}{$H_{\rm c2}^{\rm P}$}
\newcommand{\etal}{{\it et al.} }
\newcommand{\PRL}{Phys. Rev. Lett. } 
\newcommand{\PRB}{{\it Phys. Rev.} B } 
\newcommand{\JPSJ}{J. Phys. Soc. Jpn. } 
\newcommand{\Science}{{\it Science} } 
\newcommand{\Nature}{{\it Nature} } 
\newcommand{\qf}{\vec{q}_{\rm {\scriptstyle FFLO}}}
\newcommand{\qfs}{q_{\rm {\scriptstyle FFLO}}}
\renewcommand{\i}{\hspace*{0.3mm}\vec{i}\hspace*{0.6mm}}
\renewcommand{\j}{\hspace*{0.3mm}\vec{j}\hspace*{0.6mm}}
\newcommand{\Co}{CeCoIn$_5$ }
\newcommand{\Cof}{CeCoIn$_5$}
\newcommand{\va}{\vec{a}}
\newcommand{\vb}{\vec{b}}
\newcommand{\vdelta}{\vec{\delta}\hspace*{0.5mm}}
\newcommand{\vH}{\vec{H}}
\newcommand{\neel}{N$\acute{\rm e}$el }

\section{Introduction}

The possible presence of a spatially modulated state in spin polarized 
superconductors was predicted by Fulde and Ferrell~\cite{FF}, and by 
Larkin and Ovchinnikov~\cite{LO} more than 40 years ago. 
 While the Bardeen-Cooper-Schrieffer (BCS) theory assumes 
Cooper pairs with vanishing total momentum, 
the FFLO superconducting state represents a
condensate of Cooper pairs with a finite total momentum. 
 Since the FFLO state has an internal degree of freedom arising from the
reflection or inversion symmetry, a spontaneous breaking of 
the spatial symmetry occurs.  

 Although this novel superconducting state with an exotic symmetry has 
been attracting much interest, the FFLO state has not been observed in
superconductors for nearly 40 years. 
Naturally, the discovery of a new superconducting phase 
in \Co at high magnetic fields and 
low temperatures~\cite{radovan2003,PhysRevLett.91.187004} 
triggered numerous theoretical 
and experimental studies because this high-field superconducting (HFSC) 
phase is a most likely candidate for the FFLO state~\cite{matsuda2007}. 
 The recent interest on the FFLO superconductivity/superfluidity  
extends further into various related fields, such as organic 
superconductors~\cite{uji:157001,singleton2000,lortz:187002,
shinagawa:147002}, 
cold atom gases~\cite{Partridge01272006,Zwierlein01272006}, 
astrophysics, and nuclear physics \cite{casalbuoni2004}.

 Although the HFSC phase of \Co has been interpreted widely 
within the concept of the FFLO 
state,~\cite{matsuda2007,watanabe2004,capan2004,martin2005,
mitrovic2006,miclea2006,correa2007,adachi2003,ikeda:134504,ikeda:054517}
recent observations of the magnetic order in 
the HFSC phase call for a reexamination of this 
conclusion.~\cite{young2007,kenzelmann2008} 
 Neutron scattering measurements have shown that 
the wave vector of the AFM order is incommensurate 
$\Q = \Qaf \pm \qiv$ with $\Qaf = (\pi,\pi,\pi)$ and the incommensurate 
wave vector is fixed to $\qiv \sim (0.12\pi,\pm 0.12\pi,0)$ independent
of the orientation of magnetic 
field.~\cite{kenzelmann2008,kenzelmann2010}
 The AFM staggered moment $\vec{M}_{\rm AF}$ lies 
along the {\it c}-axis.~\cite{kenzelmann2008,kenzelmann2010} 
 For the magnetic field along [110] direction, the 
incommensurate wave vector is perpendicular to the magnetic field 
$\qiv \sim (0.12\pi,- 0.12\pi,0)$, while the two degenerate 
incommensurate AFM states with $\qiv \sim (0.12\pi,0.12\pi,0)$ 
and $\qiv \sim (0.12\pi,-0.12\pi,0)$ 
appear in the magnetic field along 
[100] direction.~\cite{kenzelmann2008,kenzelmann2010} 

 The magnetic order of \Co appears in the superconducting state, 
but not in the normal state.  
 This coupling between the magnetism and superconductivity 
is in sharp contrast to the other heavy fermion 
superconductors where the magnetic order is suppressed 
by superconductivity.~\cite{kitaoka2004} 
 Another important feature of \Co is the enhancement of 
HFSC phase under pressure.~\cite{miclea2006} 
This feature is also in contrast to conventional magnetic order 
in Ce-based heavy fermion systems which is suppressed 
by pressure.~\cite{kitaoka2004} 
 These unusual features indicate that 
the magnetic order of \Co is not a conventional AFM order.

 Several intriguing quantum phases have been theoretically proposed 
for the unconventional magnetic order in \Cof.~\cite{yanase_LT,
yanase_JPSJ2009,miyake2008,aperis2008,aperis2010,agterberg2009,ikedaAF} 
 Our proposal is based on the presence of AFM quantum critical point 
near the superconducting phase of \Cof~\cite{bianchi2003,ronning2005}. 
 We have shown that the AFM order occurs when the inhomogeneous 
Larkin-Ovchinnikov state is stabilized in the vicinity of 
the AFM quantum critical point.~\cite{yanase_LT,yanase_JPSJ2009} 
 The coupling between the magnetism and FFLO superconductivity 
seems to be consistent with the above-mentioned unusual feature 
that the magnetically ordered phase is confined in the 
superconducting phase in the $H$-$T$ phase 
diagram.~\cite{young2007,kenzelmann2008} 
 Another proposal has been given on the 
basis of the emergence of a pair density wave state.~\cite{
aperis2008,aperis2010,agterberg2009}

 In order to identify the HFSC phase of \Co it is highly desirable 
to investigate these proposed phases for a comparison with 
experimental results. 
 For this purpose, we investigate the AFM phases in the FFLO state. 
 We show that the multiple phases appear in the AFM-FFLO state 
when the magnetic field is applied along [100] or [010] direction. 
 We discuss the possible phase diagram of \Co 
on the basis of the comparison with experiments.

\section{Formulation}
 
An intriguing magnetic phase diagram arises from the 
degeneracy between the incommensurate wave vector 
$\qiv \sim (0.12\pi,0.12\pi,0)$ and $\qiv \sim (-0.12\pi,0.12\pi,0)$
in the magnetic field along [100] or [010] direction. 
 In order to investigate the magnetic phases, 
we formulate the phenomenological 
two component Ginzburg-Landau model. 
 We here consider the AFM order in the inhomogeneous 
Larkin-Ovchinnikov state 
for which the Ginzburg-Landau functional of the free energy is 
described as, 
\begin{eqnarray}
\label{eq:GL}
  && \hspace*{-10mm}
F(\eta_1, \eta_2)/F_0 = 
[(T/T_{\rm N}^{0} -1) + \xi_{\rm AF}^{2}(\qa - \qia)^{2}] \eta_1^2
\nonumber \\ && \hspace*{+11mm} 
+[(T/T_{\rm N}^{0} -1) + \xi_{\rm AF}^{2}(\qb - \qib)^{2}] \eta_2^2
\nonumber \\ && \hspace*{+11mm} 
+ \frac{1}{2}  (\eta_1^2 + \eta_2^2)^{2}
+ b \eta_1^2 \eta_2^2 
+ c_1 H_{\rm x} H_{\rm y} (\eta_1^2 - \eta_2^2)
\nonumber \\ && \hspace*{+11mm} 
- \frac{1}{2} \eta_1 \eta_2 \sum_{n} c_{2}(n) \delta(\qa - \qb, 2 n \qf), 
\end{eqnarray}
where $\eta_1$ and $\eta_2$ are the two component order parameters 
corresponding to the two degenerate AFM states with 
$\Q =  \Qaf \pm \qa \sim \Qaf \pm \qia$ and 
$\Q =  \Qaf \pm \qb \sim \Qaf \pm \qib$, respectively.
 The wave vector of incommensurate AFM order is assumed as 
$\qia = (0.125\pi,0.125\pi,0)$ and 
$\qib = (-0.125\pi,0.125\pi,0)$. 
 Small deviations of wave vectors from $\qia$ and $\qib$ 
due to the pinning of FFLO nodal planes are taken into account 
in eq.(\ref{eq:GL}). 
 In our study the incommensurate wave vectors $\qia$ and $\qib$ are not 
microscopically derived but assumed on the basis of the experimental 
results in Refs.~23 and 24. 
 We think that the $\qia$ and $\qib$ are pinned through nesting 
features in the band structures.

 We define the AFM staggered moment $M_{\rm AF}(\r) = (-1)^{x+y+z} M(\r)$
where $M(\r)$ is the magnetic moment perpendicular to the applied 
magnetic field at $\r = (x,y,z)$. 
 Then, the AFM staggered moment is described as, 
\begin{eqnarray}
 \label{eq:moment}
  && \hspace*{-10mm}
M_{\rm AF}(\r) = 
M_0 [\eta_1 \cos(\qa \cdot \r) + \eta_2 \cos(\qb \cdot \r)], 
\end{eqnarray}
where $M_0$ (and $F_0$) are scaling factors by which 
the Ginzburg-Landau free energy is written in the renormalized form 
(eq.(\ref{eq:GL})). 
 We take the unit of magnetic moment and energy so that $M_0 = F_0 =1$. 
 The $T_{\rm N}^{0}$ is the \neel temperature for $c_1 = c_2(n) = 0$ 
and $\xi_{\rm AF}$ is the coherence length of AFM order. 

 The phase diagram of the Ginzburg-Landau model  
considerably depends on the sign of the non-linear quartic 
coupling term $b$. 
 The single-$q$ magnetic structure $(\eta_1, \eta_2) \propto (1,0)$ 
or $(0,1)$ is stabilized by the positive coupling constant $b$,  
while the negative $b$ favors the double-$q$ magnetic structure 
$(\eta_1, \eta_2) \propto (1,\pm 1)$.

 The coupling constant $c_1$ describes the effect of magnetic 
field $\vH = (H_{\rm x},H_{\rm y},H_{\rm z})$ which may break 
the degeneracy of $\eta_1$ and $\eta_2$. 
 According to the neutron scattering measurement 
for $\vH \parallel [110]$,~\cite{kenzelmann2008} 
$c_1$ is positive in \Cof. 
 This is consistent with our theoretical analysis based on the 
Bogoliubov-de-Gennes (BdG) 
equations.~\cite{yanase_LT,yanase_JPSJ2009,yanase_JPCM} 
 When we consider the magnetic field along [100] direction, 
this term vanishes and therefore the degeneracy of $\eta_1$ 
and $\eta_2$ remains. 
 We focus on this case in this paper.

 Effects of the broken translation symmetry in the inhomogeneous 
Larkin-Ovchinnikov state are taken into account 
in the commensurate term, that is the last term of eq.(\ref{eq:GL}). 
 We define $\delta(\qa - \qb, 2 n \qf) = 1$ 
when the commensurate condition $\qa - \qb = 2 n \qf$ is satisfied 
for an integer $n$, and otherwise $\delta(\qa - \qb, 2 n \qf) = 0$. 
 The modulation vector of FFLO state is denoted as $\qf$, 
and then the order parameter of superconductivity is described as
$\Delta(\r) = \Delta_0 \sin(\qf \cdot \r)$. 
 The commensurate term describes the pinning effect of FFLO nodal planes 
for the AFM moment. 
 According to our analysis based on the BdG equation, 
the AFM moment is enhanced around the FFLO nodal planes where 
the superconducting order parameter 
vanishes.~\cite{yanase_LT,yanase_JPSJ2009,yanase_JPCM}  
 Then, we obtain the positive coupling constant $c_2(n) \geq 0$.

\section{Magnetic Phases}

 We here consider the FFLO state in which the higher harmonic 
FFLO wave vector $2N\qf$ is close to $\qia - \qib$. 
 Then, the broken translation symmetry plays an important 
role for the appearance of magnetic phases as will be shown below. 
 When the wave vector $\qia - \qib$ is not in the vicinity of 
$2n\qf$ for any integer $n$, the commensurate term can be neglected. 
Then, the single-$q$ magnetic phase with 
$(\eta_1, \eta_2) \propto (1,0)$ or $(0,1)$ is realized for 
a positive $b$, while the double-$q$ phase with 
$(\eta_1, \eta_2) \propto (1,1)$ is stabilized for a negative $b$. 
 We choose the parameter $c_2(N) =0.01$ in the 
following part and investigate the magnetic phase diagram 
for various $b$. 

 We determine the magnetic phase by minimizing the 
Ginzburg-Landau free energy (eq.(\ref{eq:GL})) with respect to 
the order parameters $\eta_1$ and $\eta_2$ and their momentum 
$\qa$ and $\qb$ for each temperature $T$ and FFLO wave vector $\qf$. 
 We assume $\eta_1 \geq \eta_2 \geq 0$ without any loss of 
the generality. 
 As for the direction of $\qf$, we assume 
$\qf = \qfs \hat{x}$ for $\vH \parallel [100]$ as 
in Refs.~27 and 35. 
 Then, the phase diagram is determined by 
the renormalized parameters $T/T_{\rm N}^{0}$ and $\xi_{\rm AF}q_0$,
where the parameter $q_0$ describes the mismatch of the higher harmonic 
FFLO wave vector $2 N \qf$ and the incommensurability along 
the $\hat{x}$-axis $\qia -\qib$. 
 We define $q_0$ as 
$\qia - \qib + 2 q_0 \hat{x} = (2 \qi + 2 q_0) \hat{x} = 2 N \qf$
with $\qi = 0.125\pi$. 
 Note that $q_0 = 0$ when the commensurate condition 
$\qia - \qib = 2 N \qf$ is satisfied. 
 Since the FFLO wave number $\qfs$ increases with increasing the 
magnetic field and/or decreasing the temperature,~\cite{matsuda2007} 
the parameter $q_0$ increases with the magnetic field, and change 
its sign at the commensurate line in the $H$-$T$ phase diagram 
on which the commensurate condition is satisfied.

 We find that three magnetic phases are stabilized. 
One is the single-$q$ phase where 
\begin{eqnarray}
 \label{eq:single-q}
  && \hspace*{-5mm}
\eta_1 > \eta_2,
\\ && \hspace*{-5mm}
 \label{eq:single-q1}
\qa = \qia + (1-x) q_0 \hat{x},
\\ && \hspace*{-5mm}
 \label{eq:single-q2}
\qb = \qib - (1+x) q_0 \hat{x}.
\end{eqnarray}
 The mirror symmetry with respect to the $x$- and $y$-axes is 
spontaneously broken in the single-$q$ phase. 
 The others are the double-$q$ phase and  double-$q$' phase
where 
\begin{eqnarray}
 \label{eq:double-q}
  && \hspace*{-10mm}
\eta_1 = \eta_2,
\\ && \hspace*{-10mm}
 \label{eq:double-q1}
\qa = \qia + q_0 \hat{x},
\\ && \hspace*{-10mm}
 \label{eq:double-q2}
\qb = \qib - q_0 \hat{x}, 
\end{eqnarray}
and 
\begin{eqnarray}
 \label{eq:double-q'}
  && \hspace*{-10mm}
\eta_1 = \eta_2,
\\ && \hspace*{-10mm}
 \label{eq:double-q'1}
\qa = \qia,
\\ && \hspace*{-10mm}
 \label{eq:double-q'2}
\qb = \qib, 
\end{eqnarray}
respectively. 
\begin{figure}[htbp]
\begin{center}
\includegraphics[width=7cm]{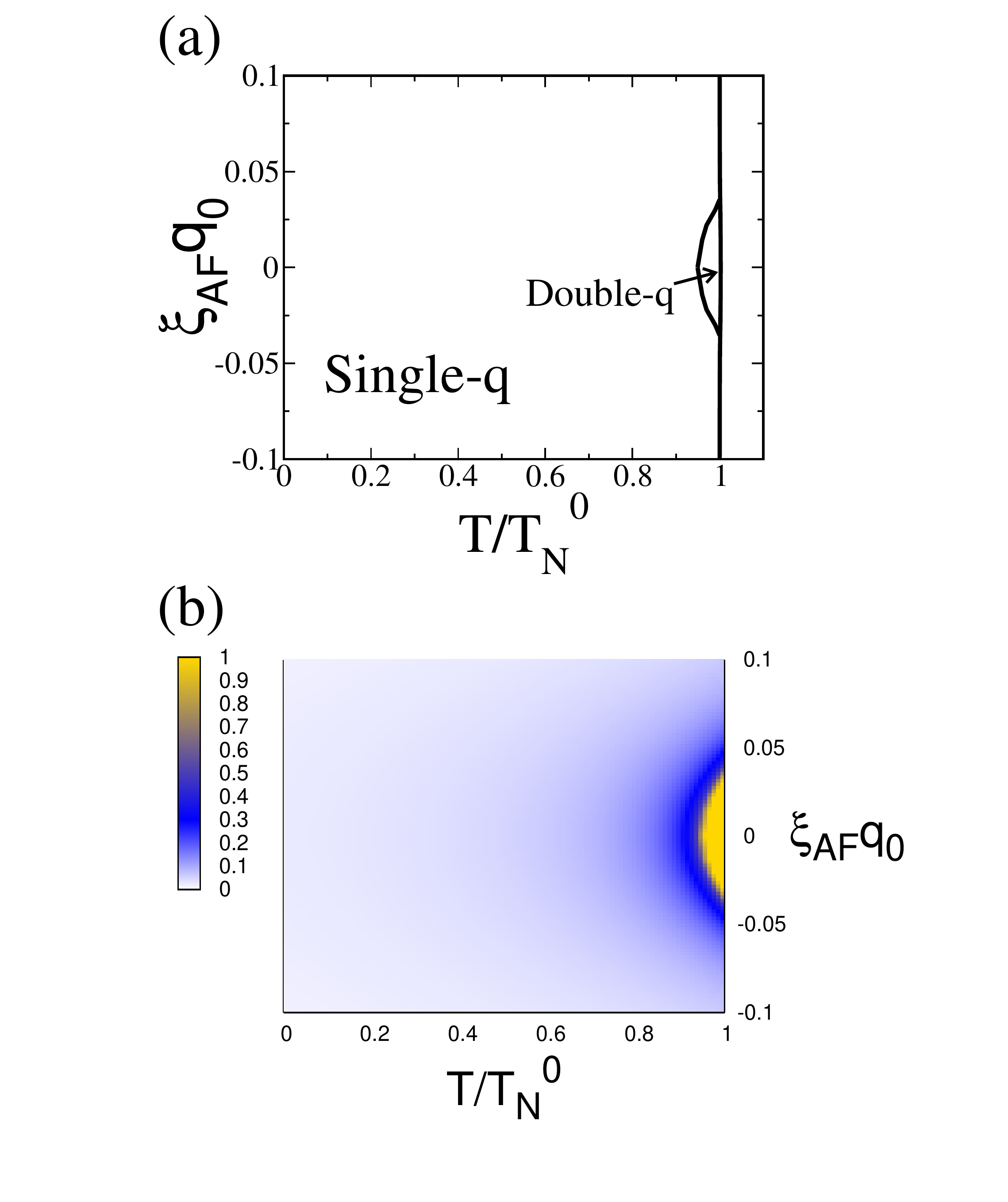}\hspace{1pc}%
\caption{(Color online)
(a) Phase diagram of the Ginzburg-Landau model with $b=0.1$
for $\xi_{\rm AF}q_0$ and the renormalized temperature 
$T/T_{\rm N}^{0}$. 
The definition of $q_0$ is given in the text. 
The $q_0$ increases with the magnetic field. 
(b) The ratio of order parameters $\eta_1/\eta_2$. 
}
\end{center} 
\end{figure}
\begin{figure}[htbp]
\begin{center}
\includegraphics[width=7cm]{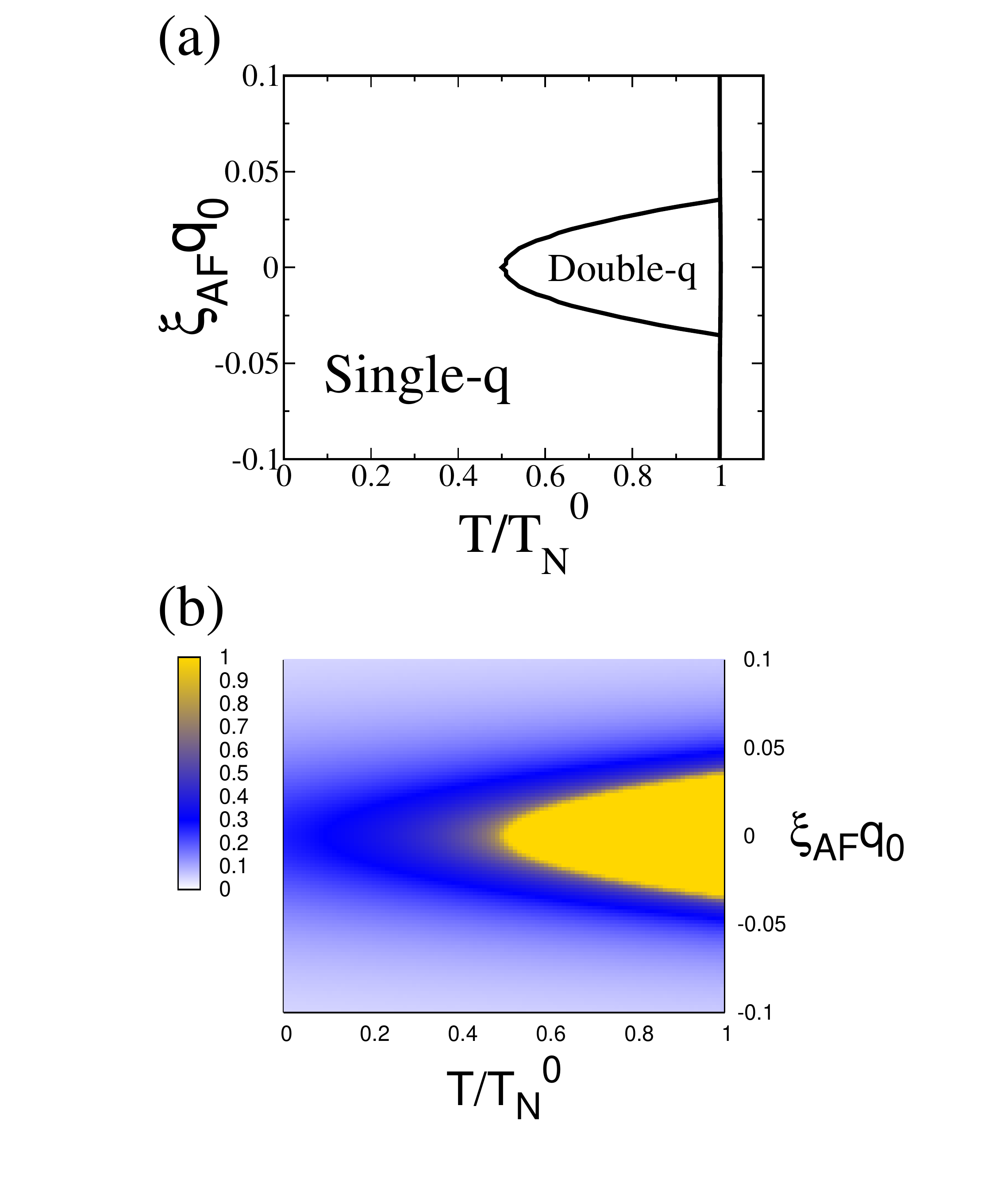}\hspace{1pc}%
\caption{(Color online)
(a) Phase diagram and (b) the ratio of order parameters $\eta_1/\eta_2$ 
for $b=0.01$. 
}
\end{center} 
\end{figure}

 The analysis of the quadratic terms in eq.(\ref{eq:GL}) 
shows that the double-$q$ phase 
is stabilized immediately below the \neel temperature for 
$(\xi_{\rm AF}q_0)^{2} \leq c_2(N)/8$, while a 
single-$q$ phase 
is stabilized for $(\xi_{\rm AF}q_0)^{2} > c_2(N)/8$. 
 The \neel temperature is 
increased around the commensurate line $\xi_{\rm AF}q_0 =0$
owing to the broken translation symmetry. 
 However, the enhancement of the \neel temperature 
$T_{\rm N} - T_{\rm N}^{0} = \frac{c_2(N)}{4}  T_{\rm N}^{0}$ 
is negligible unless the coupling constant $c_2(N)$ is large.

 Figures~1(a)-4(a) show the phase diagram obtained by the numerical 
calculation for various coupling constants $b$, while 
Figs.~1(b)-4(b) show the ratio $\eta_1/\eta_2$. 
 First, we discuss the magnetic phases for a positive $b$. 
 We see that the single-$q$ phase is stable at low temperatures 
in Figs.~1 and 2. 
 This is because the quartic term $b \eta_1^2 \eta_2^2$ 
favors the single-$q$ phase. 
 Figure~1 shows that the phase diagram is mostly covered by the 
single-$q$ phase for a large coupling constant $b \gg c_2(N)$.  
 Then, the double-$q$ phase appears around the \neel temperature 
when $|\xi_{\rm AF}q_0| \leq \sqrt{c_2(N)/8}$, but it disappears 
by decreasing the temperature.  
 The double-$q$ phase is stabilized in the intermediate temperature 
region $T \sim 0.5 T_{\rm N}$ for a small and positive 
$b = c_2(N)$ (Fig.~2). 
 Figures~1(b) and 2(b) show the ratio $\eta_2/\eta_1$ 
decreases in the single-$q$ phase 
with decreasing temperature $T/T_{N}^0$ 
and/or increasing the mismatch $|\xi_{\rm AF}q_0|$. 
 The phase transition between the single-$q$ and double-$q$ phases 
is second order.

 Next, we discuss the magnetic phases for a negative coupling constant $b$. 
Figures~3(a) and 4(a) show that the double-$q$ phase is stable at low 
temperatures for $|\xi_{\rm AF}q_0| \leq \sqrt{c_2(N)/4}$ while 
the double-$q$' phase is stabilized 
when the mismatch is larger than the critical value 
$|\xi_{\rm AF}q_0| > \sqrt{c_2(N)/4}$. 
 The single-$q$ phase appears around the \neel temperature 
near $|\xi_{\rm AF}q_0| = \sqrt{c_2(N)/4}$.   
 It is shown that the phase diagram is mostly covered 
by the double-$q$ and double-$q$' phases. 
 The phase transition is first order between the double-$q$' phase 
and the other phases, while that from the double-$q$ phase to 
the single-$q$ phase is still second order. 

\begin{figure}[htbp]
\begin{center}
\includegraphics[width=7cm]{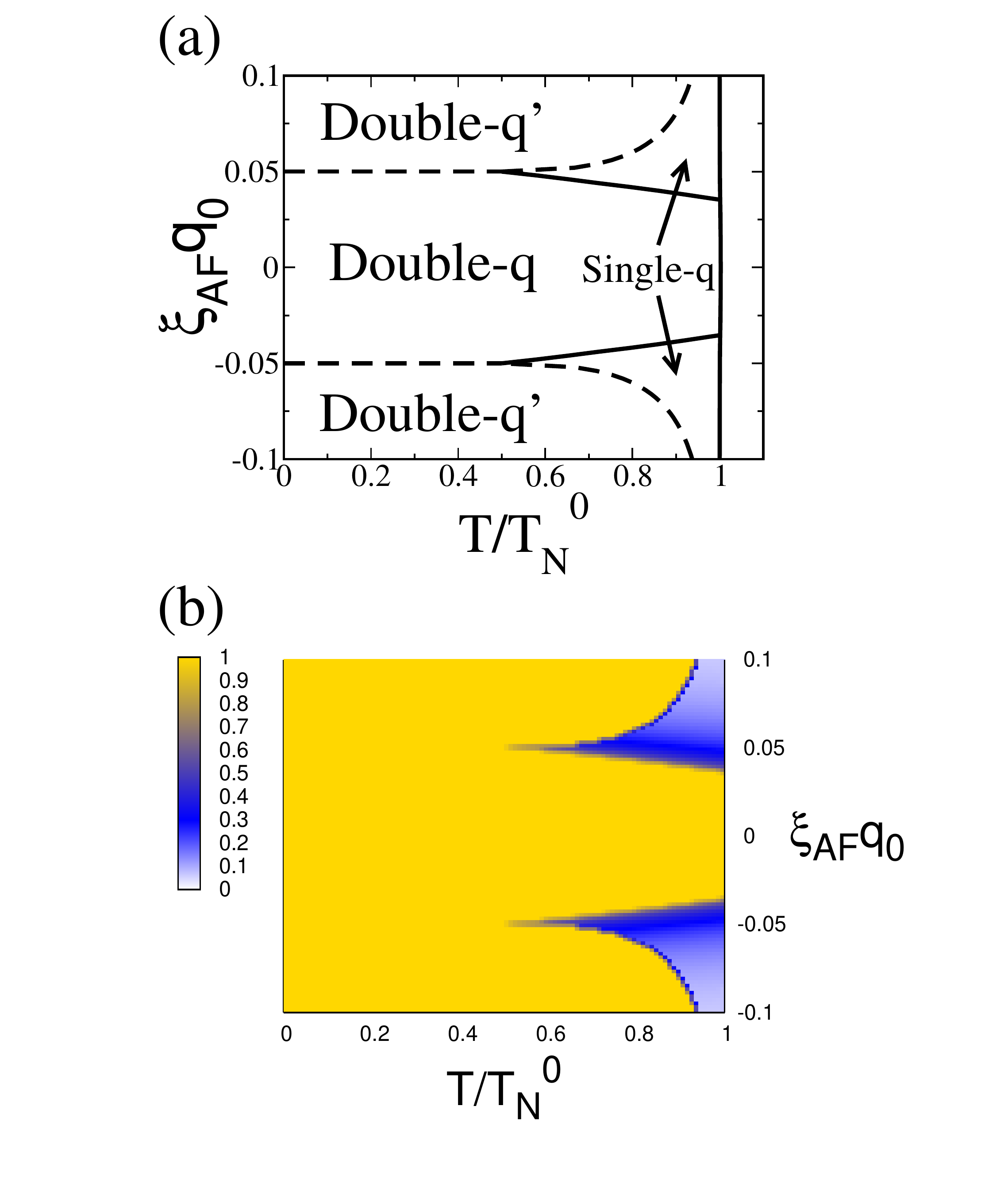}\hspace{1pc}%
\caption{(Color online)
(a) Phase diagram and (b) the ratio of order parameters $\eta_1/\eta_2$ 
for $b=-0.01$. 
The dashed lines in (a) show the first order phase transition. 
}
\end{center} 
\end{figure}

\begin{figure}[htbp]
\begin{center}
\includegraphics[width=7cm]{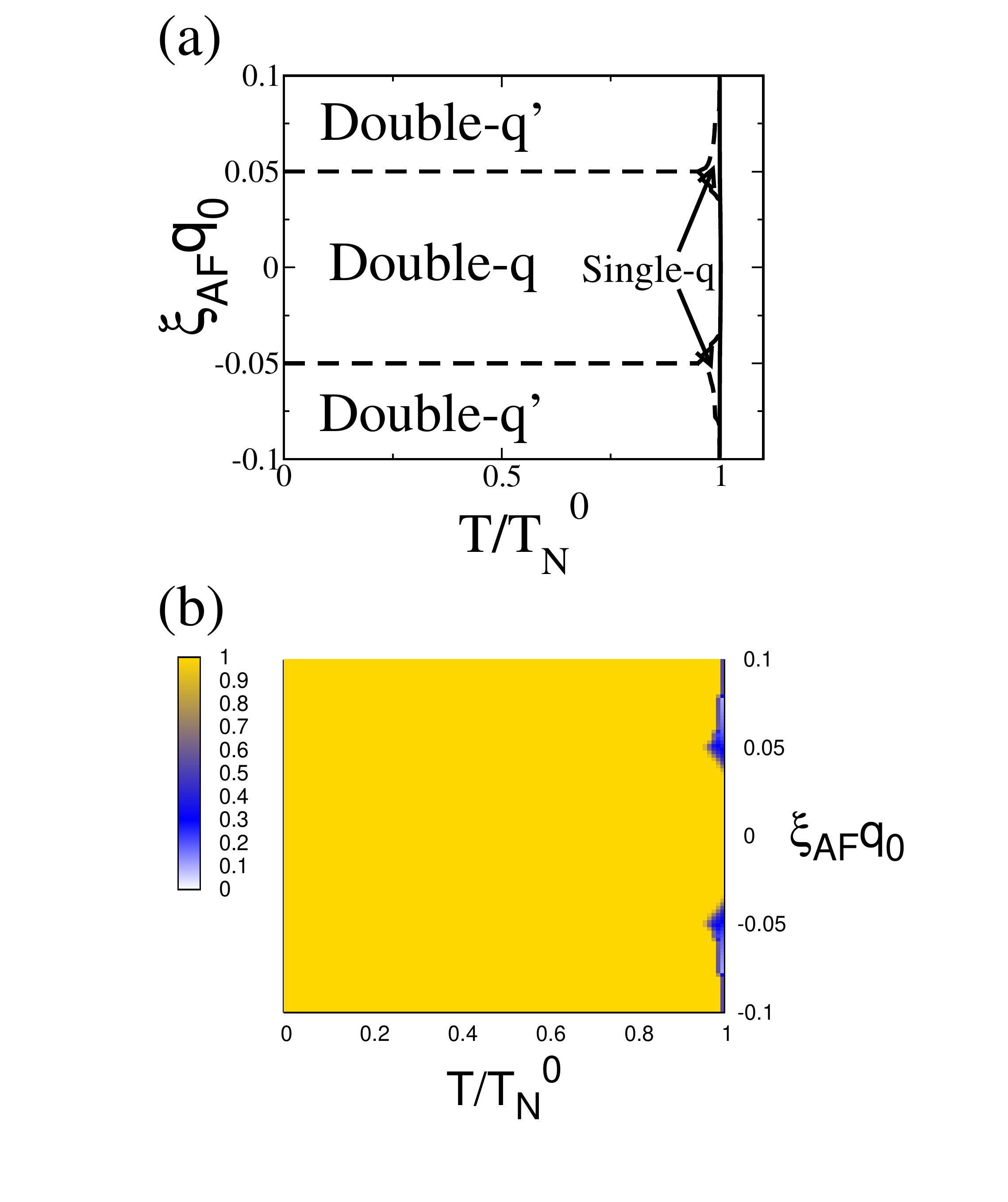}\hspace{1pc}%
\caption{(Color online)
(a) Phase diagram and (b) the ratio of order parameters $\eta_1/\eta_2$ 
for $b=-0.1$. 
The dashed lines in (a) show the first order phase transition. 
}
\end{center} 
\end{figure}

\begin{figure}[ht]
\begin{center}
\includegraphics[width=8cm]{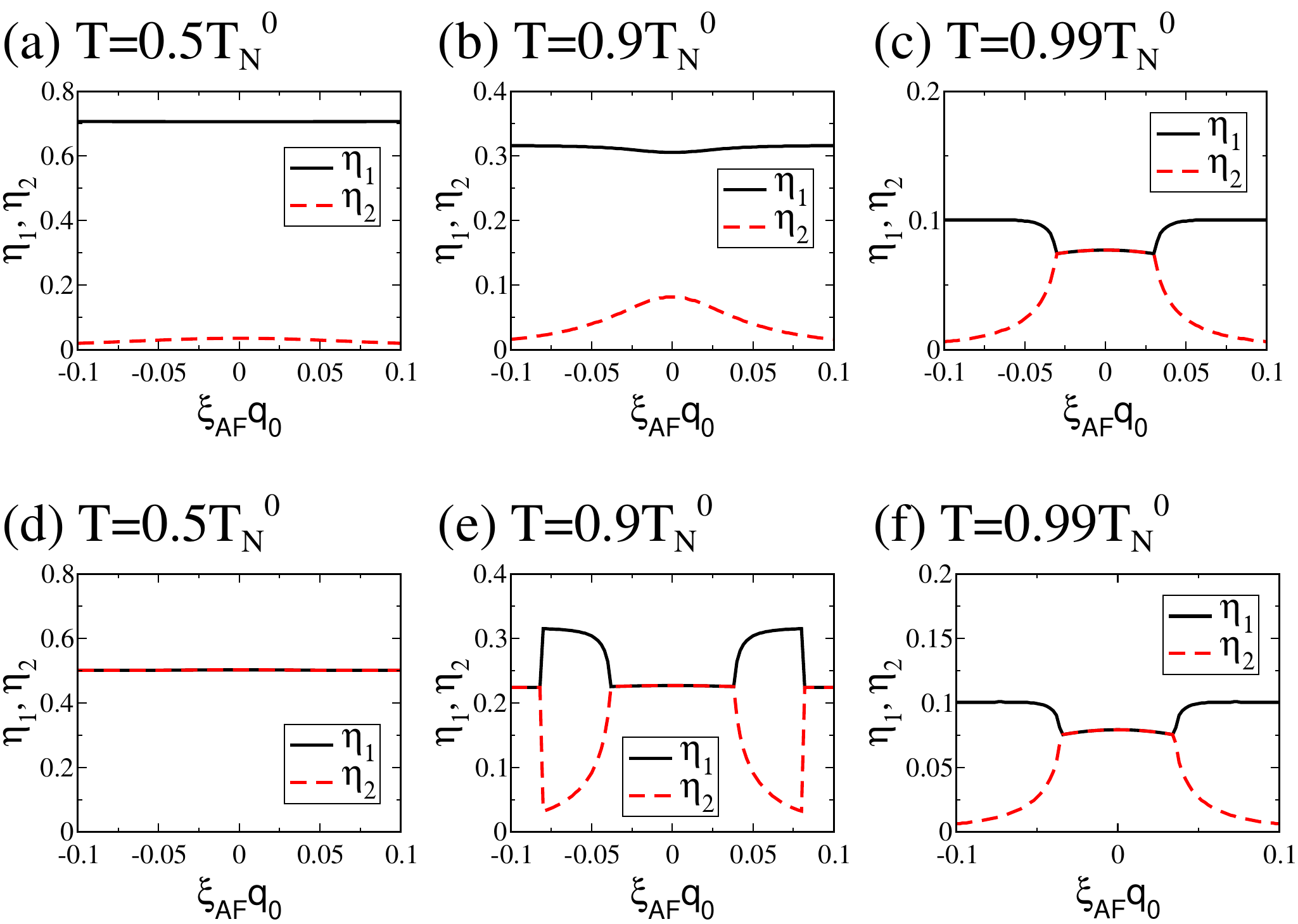}\hspace{1pc}%
\caption{(Color online)
$\xi_{\rm AF}q_0$ dependences of the order parameters 
$\eta_1$ and $\eta_2$ for $b=0.1$ (Figs.~5(a)-5(c)) 
and $b=-0.01$ (Figs.~5(d)-5(f)). 
$T=0.5T_{\rm N}^{0}$ in Figs.~5(a) and 5(d), 
$T=0.9T_{\rm N}^{0}$ in Figs.~5(b) and 5(e), and 
$T=0.99T_{\rm N}^{0}$ in Figs.~5(c) and 5(f), 
respectively. 
}
\end{center} 
\end{figure}

 In order to understand the magnetic phase diagram more clearly, 
we plot the $\xi_{\rm AF}q_0$ dependence of order parameters 
for $b=0.1$ in Figs.~5(a)-5(c) and 
that for $b=-0.01$ in Figs.~5(d)-5(f). 
 It is shown that the magnetic phases are almost independent of $b$ 
around the \neel temperature (Figs.~5(c) and 5(f)), while 
the single-$q$ phase (double-$q$ or double-$q$' phase) 
is stable at low temperatures for the positive (negative) coupling 
constant $b$.  
 Fig.~5(e) shows a discontinuous jump of order parameters 
at the first order transition from the single-$q$ phase 
to the double-$q$' phase.

\section{High Field Phase Diagram of \Cof}

 We discuss the possible AFM-FFLO state of \Co on the basis of the 
Ginzburg-Landau theory given in this paper.
 For this purpose, we show the schematic phase diagram of \Co 
for the magnetic field and temperature in Fig.~6. 

 The maximum amplitude of the FFLO modulation vector $\qfs$ is 
approximately the inverse coherence length $\qfs \sim 1/\xi$.  
 According to the experimental estimation of the coherence length of 
the superconductivity $\xi$,~\cite{miclea2006} 
the minimum number of $N$ which satisfies the commensurate condition 
$\qia - \qib = 2 N \qf$ is approximately obtained as $N \sim 4$. 
 Then, the commensurate condition is satisfied in the FFLO state 
on a sequence of commensurate lines (dashed lines in Fig.~6) with 
$N = 4, 5, 6, 7 ....$.

\begin{figure}[ht]
\begin{center}
\includegraphics[width=5.5cm]{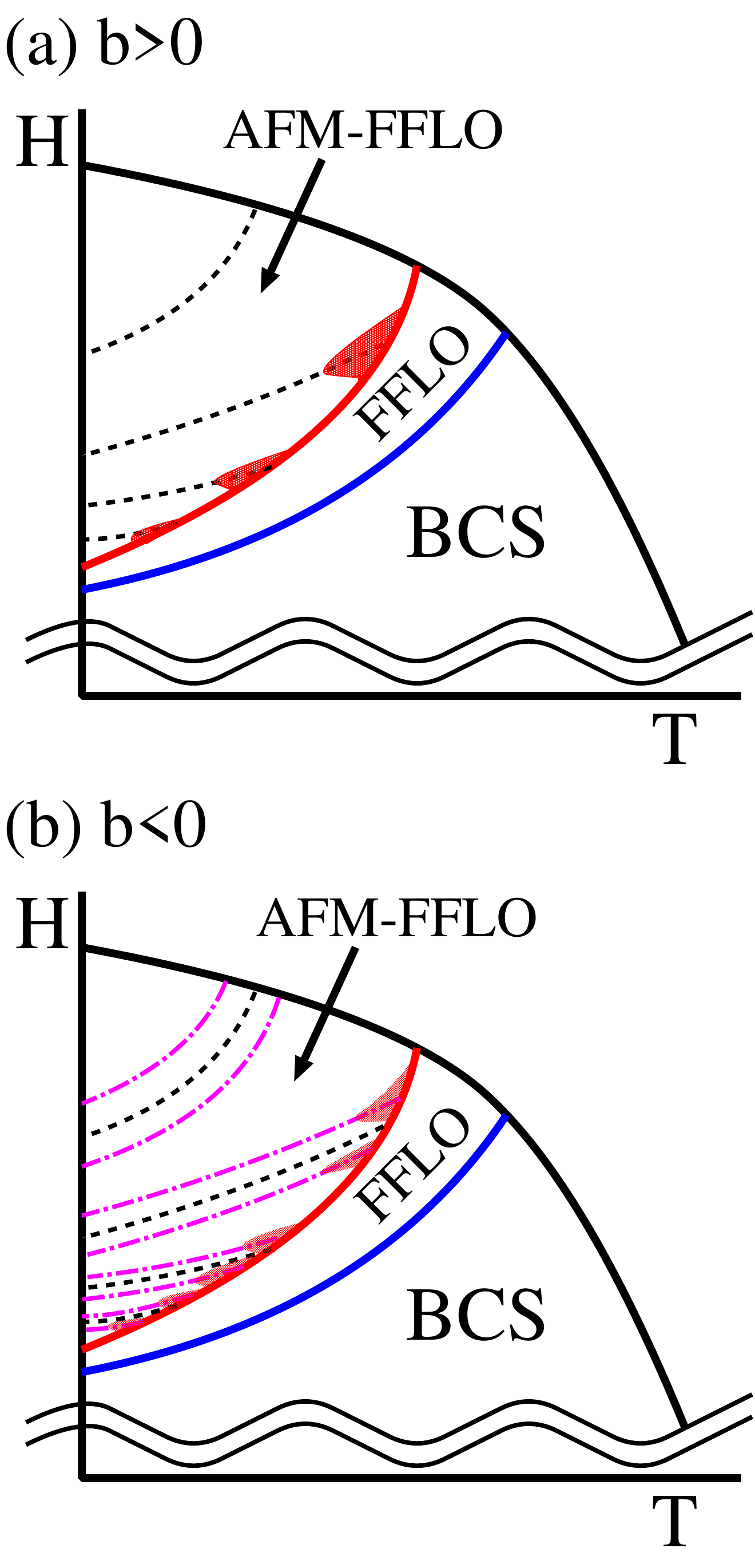}\hspace{1pc}%
\caption{(Color online)
Schematic phase diagram of \Co for the magnetic field along 
[100] direction. (a) $b > 0$ and (b) $b < 0$. 
``BCS'', ``FFLO'', and ``AFM-FFLO'' states are shown in the figure. 
The dashed lines show the commensurate lines where 
the commensurate condition is satisfied. 
 The shaded region shows the double-$q$ phase in (a), while 
it shows the single-$q$ phase in (b). 
 The dash-dotted lines in (b) show the first order 
phase transition lines between the double-$q$ phase and 
double-$q$' phase. 
}
\end{center} 
\end{figure}

 The magnetic phase diagram in the FFLO state is quite different 
between the positive and negative coupling constants $b$, as shown in 
Figs.~6(a) and 6(b). 
 Figure~6(a) shows the phase diagram for a positive $b$. 
 We see that the AFM-FFLO state is mostly the single-$q$ phase 
while the double-$q$ phase is stabilized 
around the AFM transition line near the commensurate lines 
(shaded area of Fig.~6(a)). 
 Note that the double-$q$ phase does not appear 
around the first order normal-to-FFLO transition line 
at which the AFM moment as well as the superconducting order parameter 
are discontinuous. 
 Since it is expected that the coupling constant $c_2(N)$ decreases 
with increasing $N$, 
the double-$q$ phase is suppressed with decreasing the magnetic field. 

 Figure~6(b) shows a schematic phase diagram for a negative 
coupling constant $b$, where the double-$q$ and double-$q$' 
phases are stable at low temperatures. 
 In contrast to Fig.~6(a), a sequence of 
discontinuous first order transition lines appear in the 
AFM-FFLO state. These first order transitions occur due to the 
pinning of AFM moment on the FFLO nodal planes. 
 This is a direct consequence of the broken translation symmetry 
in the inhomogeneous Larkin-Ovchinnikov state. 
 However, these first order transitions have not been observed in the 
experiments of \Cof. 
 This indicates that the coupling constant $b$ is positive in \Cof. 
 Thus, the phase diagram in Fig.~6(a) is more likely realized in \Co 
than Fig.~6(b). 
 Since the second order phase transition shown in Fig.~2 has not been 
observed in the magnetic phase of \Cof, The coupling constant $b$ 
should be much larger than $c_2(n)$. 
 Then, the AFM-FFLO state is mostly covered by the 
single-$q$ phase where the mirror symmetry with respect to the 
{\it x}- and {\it y}-axis is broken. 
 It is desirable that the phase diagram of \Co will be investigated 
by the intensive experiments in the magnetic field along [100] axis.

\begin{figure}[ht]
\begin{center}
\includegraphics[width=6cm]{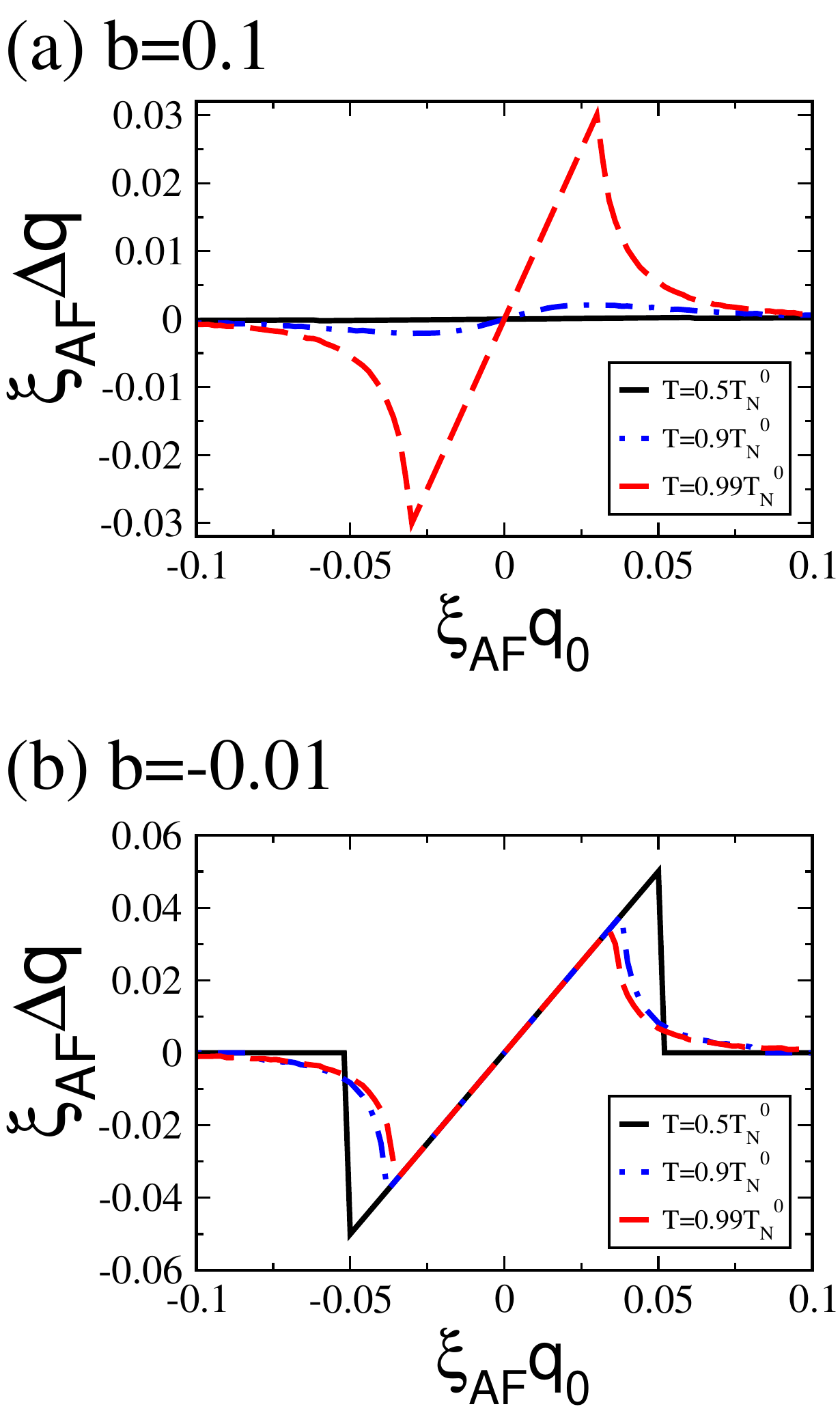}\hspace{1pc}%
\caption{(Color online)
The shift of the position of main Bragg peak from $\qia$. 
We plot $\xi_{\rm AF} \Delta q$ for $T=0.5T_{\rm N}^{0}$, 
$T=0.9T_{\rm N}^{0}$, and $T=0.99T_{\rm N}^{0}$, 
where $\Delta q$ is defined by $\Delta q \hat{x} = \qa - \qia$. 
}
\end{center} 
\end{figure}

 The phase diagram in Fig.~6(a) is supported by the 
nuclear magnetic resonance (NMR) and 
neutron scattering  measurements for \Cof. 
 We first discuss the neutron scattering measurement.
 According to the results obtained 
by Kenzelmann {\it et al.,}, the position of Bragg peaks does not 
change with increasing the magnetic field along [100] 
direction.~\cite{kenzelmann2010} 
 The phase diagram for a negative $b$ (Figs.~3 and 4) is incompatible 
with this experimental result. 
Figure~7 shows the $\xi_{\rm AF}q_0$ dependence 
of the shift of main Bragg peak $\Delta q \hat{x} = \qa - \qia$. 
For $b < 0$, Fig.~7(b) shows the shift $\xi_{\rm AF}\Delta q > 0.05$ 
at low temperatures. When we assume $\xi_{\rm AF} =3$, 
$\Delta q \sim 0.006 \pi$ near the first order transition lines. 
 This shift is larger than the experimental error,~\cite{kenzelmann2010} 
and therefore can be observed if it would occur. 
 However, the experimental result has not shown the shift of Bragg peaks. 

 On the other hand, the neutron scattering measurement is consistent 
with our results for the positive $b$. 
 Fig.~7(a) shows a pronounced shift $\xi_{\rm AF}\Delta q $ up to $ 0.03$
near the \neel temperature, but the shift rapidly decreases with 
decreasing the temperature. 
 Since the magnetic moment is tiny near the AFM transition, 
it is difficult to experimentally observe of the shift of Bragg peaks 
around $T_{\rm N}$. 
 Thus, our results in Fig.~7(a) are consistent with the experimental 
observation.~\cite{kenzelmann2010}

 We here comment on the effect of domain structure 
in the single-$q$ phase. 
 For the magnetic field along [100] direction 
the elastic Bragg peaks appear at $\Q = \Qaf \pm \qia$ 
as well as at its symmetric point 
$\Q = \Qaf \pm \qib$.~\cite{kenzelmann2008,kenzelmann2010} 
 This result seems to be incompatible with the phase diagram 
for $b>0$, because the intensity of Bragg peaks at 
$\Q = \Qaf \pm \qib$ is very weak in the single-$q$ phase with 
$|\eta_1| \gg | \eta_2|$. 
 However, the observed four Bragg peaks are consistent with 
the single-$q$ phase by 
taking into account the domain formation of 
two degenerate single-$q$ states, 
i.e. $ |\eta_1| > | \eta_2| $ and $ |\eta_1| < | \eta_2| $. 

 The phase diagram in Fig.~6(a) is also supported by the NMR measurements 
which can distinguish the single-$q$ phase from the double-$q$ phase. 
 When the hyperfine coupling has the dipolar symmetry, 
the NMR spectrum at the In(2b) site shows the distribution 
function of the internal field arising from the AFM staggered 
moment.~\cite{curro2009} 
 The experimental results show the double peak structure of 
NMR spectrum at the In(2b) site.~\cite{young2007,mitrovic2006,mitrovic2008,
koutroulakis2010,kumagaiprivate} 
 According to our analysis which will be published elsewhere, 
these results are consistent with the single-$q$ phase, 
but not with the double-$q$ phase. 
 Thus, both neutron scattering and NMR measurements support the 
positive coupling constant $b$ (Figs.~1 and 6(a)) 
and rule out the negative $b$ (Figs.~3, 4 and 6(b)).

 When we take into account the broken translation symmetry arising from 
the vortex lattice, other commensurate lines can appear in the 
phase diagram and stabilize the double-$q$ phase around $T=T_{\rm N}$. 
 However, it is expected that the effect of vortex lattice 
on the magnetic order is smaller than that of the FFLO nodal planes 
when the Maki parameter is large enough to make the lattice spacing 
of vortices much larger than the coherence length.

\section{Summary}

 We investigated the incommensurate AFM order in the 
FFLO superconducting state 
with a particular interest on the possible coexistence of 
the AFM order and FFLO superconductivity in the HFSC phase of \Cof. 
 Assuming the incommensurate AFM order with $\qiv \parallel [110]$ or 
$\qiv \parallel [1\bar{1}0]$, we examined the magnetic phase 
in the magnetic field along $[100]$ direction. 
 We find that the multiple AFM phases appear in the $H$-$T$ 
phase diagram owing to the broken translation symmetry in the FFLO 
superconducting state. 
 The comparison between the Ginzburg-Landau theory and experimental 
results shows that the AFM-FFLO state is mostly covered by 
the single-$q$ phase in which the mirror symmetry with respect 
to the $x$ and $y$ axes is broken.

\section*{Acknowledgements}

 The authors are grateful to D.F. Agterberg, S. Gerber, R. Ikeda, 
M. Kenzelmann, K. Kumagai, K. Machida, Y. Matsuda,  V. F. Mitrovi\'c, 
and H. Tsunetsugu 
for fruitful discussions. 
 This work was supported by 
a Grant-in-Aid for Scientific Research on Innovative Areas
``Heavy Electrons'' (No. 21102506) from MEXT, Japan. 
 It was also supported by a Grant-in-Aid for 
Young Scientists (B) (No. 20740187) from JSPS. 
 Numerical computation in this work was carried out 
at the Yukawa Institute Computer Facility. 
YY is grateful for the hospitality of the Pauli Center of ETH Zurich. 
This work was also supported by the Swiss Nationalfonds 
and the NCCR MaNEP.

\bibliographystyle{jpsj}
\bibliography{FFLO_GL}

\end{document}